\documentclass[a4paper]{article}
\usepackage{CJK}
\usepackage{geometry}
\usepackage{epsfig}
\usepackage{graphicx}
\usepackage{threeparttable}
\usepackage{multirow}
\usepackage{hyperref}
\usepackage{color}
\begin{document}
\title{Massive Neutral Gas Outflow in Reddened Quasar SDSS J072910.34+333634.3}
\author{Tong Liu$^{1}$, Luming Sun$^{1}$, Xiang Pan$^{1}$, and Hongyan Zhou$^{1,2}$\\ \small{$^{1}University\ of\ Science\ and\ Technology\ of\ China,\ Hefei,\ 230026,\ China$}\\ \small{$^{2}Polar\ Research\ Institute\ of\ China,\ Shanghai,\ 200136,\ China$}}
\date{}
\maketitle
\LARGE \centerline {Abstract}

\normalsize \ \\SDSS J072910.34+333634.3 is a reddened quasar at z=0.96. The archivel Keck/ESI spectrum and our new P200/TripleSpec spectrum reveal an absorption line system in He I*, Ca II H\&K and Na ID. The absorption line system has a width of $\sim$600 km/s and a blueshift velocity of $\sim$800 km/s relative to the core of narrow emission lines, indicating an outflow. Using the Ca II doublet, we determined that the outflowing gas covers $\sim$70\% of the continuum. On the other hand, the HST/ACS image which taken in rest-frame 4130 \AA\ show that the fraction of the quasar emission in ESI aperture was $<$40\%. We thus conclude that the absorbing gas covers a significant fraction of extended starlight emission, and the best-estimated fraction of $\sim$50\% yields a lower limit of the crosssectional area of the outflowing gas to be $>$8 kpc$^2$. The strong Na\&Ca absorption suggests that the absorbing gas is thick and mostly neutral, which is also supported by dust extinction $A_V\sim3$. Using the best-estimated hydrogen column density $N_H\sim3\times10^{22}$ cm$^{-2}$, the total mass of the outflowing gas is $>3\times10^9 M_\odot$. The outflow is likely to be driven by AGN because of the $\sim$800 km/s blueshift velocity, suggesting SDSS J072910.34+333634.3 is undergoing one of the most violent AGN feedback we have seen. In the future, one can find more such massive neutral gas outflows in other reddened quasars using similar method, and this can shed new light on the study of AGN feedback.


\section{Introduction}

The evolutionary scenario between ultraluminous infrared galaxy(ULIRG) and quasar ( Sanders et al. 1988; Hopkins et al. 2009) said that gas rich mergers may be the cause for major starbursts, formation of elliptical galaxies, and take responsibility for formation and growth of supermassive black holes (SMBH). The merger will first generate a ULIRG with heavy dust, which will later gradually disperse via galactic outflows, photonic sublimation, etc., evolving into reddened quasars and finally fully exposed ones. The tight SMBH-spheriod mass relation (e.g., Ferrarese \& Merritt 2000) suggests a coevolution between them. To explain this phenomenon, suppression of growth considering both the SMBH and spheroidal componet due to winds or outflows are invoked (e.g., Di Matteo et al. 2005; Veilleux et al. 2005). These winds or outflows are said to negatively impact star formation in the merger remnants, thus is called ``negative mechanical feedback¡¯¡¯.


Finding observational evidence of such feedback processes in action is one of the main challenges of current extragalactic astronomy. While there are many researches demonstrating the initial and final states of this process, observation on direct witness of the ongoing process is still rare(e.g.Sturm et al. 2011; Spoon et al. 2013, hereafter S13; Cicone et al. 2014, Hereafter C14). To qualify such an event, we have to observe an outflow:\\ a) large enough to the galactic scale, in order to drive the negative feedback on the star formation, and\\ b) possess a large enough speed to rule out the option of starburst-driven outflow.

In this paper, we report such an example, SDSS J072910.34+333634.3 (hereafter J0729 for short), which is identified as a quasar because of the high luminosity (absolute V=-25.3), existence of broad emission lines and X-ray detection.
A redshift of z=0.957 is measured from the core of the narrow emission line.
HST/ACS image reveals a remarkable core and a spatial extensive host galaxy.
By comparing the optical to infrared SED of J0729 with the composite quasar SED, we found that the quasar is heavily dusty with $A_V\sim3$.
By analysing archival Keck/ESI spectrum taken on December 30, 2000, we identified a z=0.954 absorption line system in CaII H\&K doublets and in He I* $\lambda$3889.
We thus observed the quasar using P200/TripleSpec on December 2, 2015.
The newly obtained near-infrared spectrum verified this absorption line system by revealing He I* $\lambda$10830 and Na ID absorption lines in the same redshift.
The existence of He I* absorption lines, which requires a large quantity of ionizing photon Q(He), unambiguously show that this absorption line system is directly related with the quasar.
Therefore the offset between the emission line redshift and absorption line redshift indicates that the absorbing gas is moving away from the quasar with a velocity of $\sim$800 km/s, and hence represents an outflow.
By analysing ESI and TripleSpec spectra and ACS image, we concluded the following:
\\1. The absorbing gas covers $\sim$50\% of starlight, thus is galactic scale with crosssectional area $>$8 kpc$^2$ (Section 3.1);
\\2. The absorbing gas is thick with total H column density of $\sim3\times10^{22}$ cm$^{-2}$, and thus the total mass of the absorbing gas is $>3\times10^9 M_\odot$ (Section 3.2 and 3.3).
\\This outflow is massive enough comparing to $(2-4.5)\times 10^{8} M_\odot$ in S13 and no more than $10^{6.3}\sim10^{9.5}$ in C14.
Also, the previous discoveries are more based on molecular lines, rather than atomic as what we do here.
Taking the heavy reddening and everything into consider, we speculate this object as a young core merger in the middle of gas-and-dust dispersion, where the outflow serves as a trigger in the negative mechanical feedback.
We organize this paper as follows: In chapter \ref{reduction}, we present the reduction of raw data on this object; in chapter \ref{datanalysis}, we analyse the data and calculate useful physical parameters; and in chapter \ref{discussion} we verify the qualitative results, identify the origin of the outflow and discuss physical process on this object.

\section{Observation and Data reduction}
\label{reduction}
\subsection{SED}
\normalsize In order to study its properties, we first collected its multiband photometric data, and made a spectral energy distribution (SED), which is listed in table \ref{tabsed}. We searched the original data from the database of the telescopes (please listed the names of the databases and give the links in note), usually in the form of magnitude, and used the formula
\begin{equation}
\log(\nu F_{\nu})=\frac{-M}{2.5} +\log{zeropt}+\log{\nu}
\end{equation}
where zeropt is the magnitude zeropoint, to derive its flux.\\
The reduced data are listed in table \ref{tabsed}, and the corresponding figure is plotted in figure \ref{sed-fitting}. From the radio-FIR part of the SED, we can see that this is a radio-quiet quasar. Thus, we compare them with the composite SED of radio-quiet quasars, and plot them in the same figure, where the composite SED is verically moved by a factor of 11.56.\\

\begin{table}[!htbp]
\begin{tabular}{|l|l|l|l|}
\hline
band & $\ \log \nu $\ & $\ \log \nu f_{\nu} $\ & uncertainty \\
\hline
1.4GHz & 9.438 & 6.561 & 0.084 \\
\hline
20cm(VLA) & 9.468 & 6.540 & NaN \\
\hline
3.6cm(VLA) & 10.212 & 6.665 & 0.039 \\
\hline
Spitzer MIPS $S_{160}$ & 12.564 & 10.750 & NaN \\
\hline
Spitzer MIPS $S_{70}$ & 12.923 & 10.808 & NaN \\
\hline
Spitzer MIPS $S_{24}$ & 13.388 & 11.207 & 0.101 \\
\hline
WISE W4 & 13.426 & 11.215 & 0.039 \\
\hline
WISE W3 & 13.689 & 11.407 & 0.01 \\
\hline
WISE W2 & 14.106 & 11.517 & 0.009 \\
\hline
WISE W1 & 14.237 & 11.332 & 0.009 \\
\hline
2MASS K & 14.434 & 11.159 & 0.036 \\
\hline
2MASS H & 14.548 & 11.092 & 0.071 \\
\hline
2MASS J & 14.677 & 11.09 & 0.0408 \\
\hline
SDSS z & 14.808 & 10.712 & 0.012 \\
\hline
F814W(HST)AB & 14.848 & 10.056 & 0.028 \\
\hline
SDSS i & 14.886 & 10.366 & 0.008 \\
\hline
SDSS r & 14.974 & 10.006 & 0.016 \\
\hline
SDSS g & 15.090 & 9.634 & 0.028 \\
\hline
SDSS u & 15.219 & 9.619 & 0.076 \\
\hline
NUV(GALEX)AB & 15.402 & 9.565 & 0.121 \\
\hline
0.2-10keV Chandra& 18.666 & 9.482 & NaN \\
\hline
\end{tabular}\\
\caption{SED data points for J0729.NaN stands for those whose uncertainty is not given by the observer.}
\label{tabsed}
\end{table}

\subsection{HST Image}
\normalsize HST-ACS obtained two spatial resolved images of J0729 on Oct 21 and 22, 2005, using filter F814W and F475W, whose central wavelengths are 4745\AA \ and 8061\AA \ respectively. The exposure time of F815W is 4650s, and that of F475W is 4760s. We utilized its pipeline-reduced data, and directly used them for further analysis. During the same day`s observation, HST also obtained one picture for each filter of STAR-072904.59+333636.40, each having a exposure time of 120s. We also downloaded the pipeline image of this object, whose usage will be stated in Chapter 3.

\subsection{ESI spectrum}
The Keck ESI optical long slit spectrum of J0729 was obtained on 2000 December 30 using the echelle mode. The width of the slit is 0.75 arcsec and the instrumental dispersion is $\sigma\sim22$ km s$^{-1}$. The exposure time was 15 minutes. We downloaded the original data from Keck Observatory Archive (KOA) and reduced the data with the IDL program package XIDL (http://www2.keck.hawaii.edu/inst/esi/ESIRedux/indx.html)  via standard routine. We extracted one-dimensional spectrum from an aperture of 3 arcsec in the spatial direction and in this way we obtained the final spectrum from a rectangular region with $3\times0.75$ arcsec$^2$.

\subsection{Triplespec spectrum}
A near-infrared spectrum of J0729 is acquired with TripleSpec (Wilson et al. 2004) on the 200-inch Hale Telescope at Palomar Observatory on December 2, 2015. With a slit of 1.0" X 30", 8 X 180 s exposures were taken in an A-B-B-A dithering mode. We observed a nearby A0V star quasi-simultaneously for telluric correction and flux calibration. The data were then reduced with the Spextool package (Cushing et al. 2004) with standard configurations. A 1-d spectrum spanning a wavelength of 0.97¨C2.46 micron at a resoltion of R$\sim$2500 is then obtained. Median SNR of the spectrum is $\sim$7, the deep Na ID and He I* $\lambda$10830 absorption troughs are clearly seen in it.
\section{Data Analysis}
\label{datanalysis}

\normalsize After we obtain all the original data from section \ref{reduction}, we are now ready to analyse the physical conditions of this object, starting from absorption lines.
From chapter \ref{reduction}, we obtained and reduced ESI and TripleSpec spectrum. In these spectrum there are quite a few absorption lines that can shed light on important properties of this object. So, at the beginning of this section, we first normalize, show and identify all the absorption lines we might use throughout the passage.

\begin{figure}[!htbp]
\centering
\includegraphics[width=\linewidth]{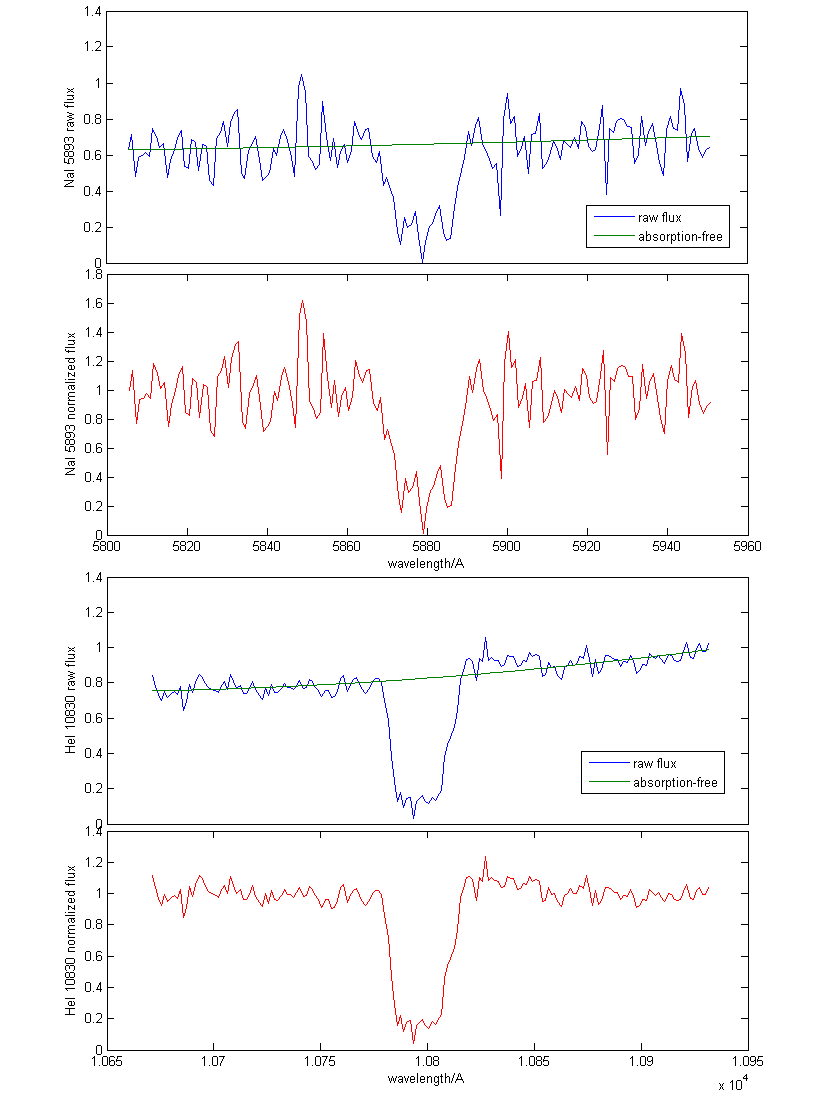}
\caption{Spectra normalization of NaI 5893(top) and HeI* 10830(bottom). In all figures, blue lines stand for raw spectrum, green stand for absorption-free fit, and red stand for normalized spectrum.}
\label{absorptionfree}
\end{figure}

For analysing the absorption line system, we have to first obtain the absorption free spectrum, and the normalized absorption spectrum by dividing the original spectrum by it.
The absorption free spectrum in the Ca II and He I* $\lambda$3889 region was obtained using pair-matching method. This method is based on the similarity of continua and emission line profile between quasars with and without absorption.
We did the pair-matching process following Liu15, who developed the method to analyze He I* $\lambda$3889 BALs.
The detailed process can be found in Liu15 and we summarized it as below.
We looped over the library of unabsorbed spectra which is generated by Liu15, and fit the spectrum of J0729 with these unabsorbed spectra by multiplying a two-order polynomial, and selected the best-fittings via $\chi^2$ to produce the final absorption-free spectrum.
The Na ID and He I* $\lambda$10830 apsorption free spectra is obtained by the fitting the nearest absorption-free part of the spectrum on both sides of the absorption line with polynomials.
Then, we divide the raw flux by the absorption free fit, and obtained the normalized spectra.
The raw as well as normalized spectra are shown in figure \ref{absorptionfree}.\\

\begin{figure}[!htbp]
\centering
\includegraphics{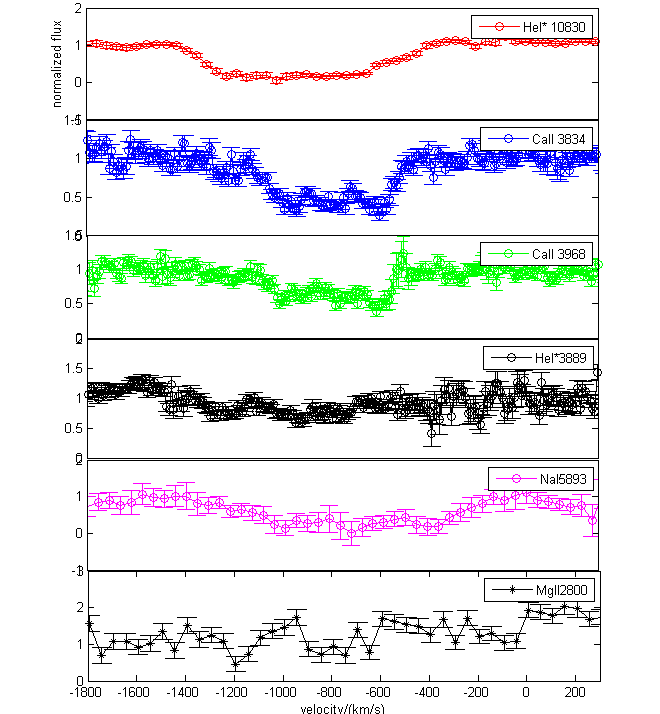}
\caption{Absorption Lines. From top to bottom, the lines are respectively HeI* 10830, CaII 3934, 3968, HeI 3889, NaID, and MgII 2800. The typo of HeI 3889 in the figure will be corrected in later editions.}
\label{lines}
\end{figure}
\normalsize The absorption lines available are shown in \ref{lines}, where the velocity is based on the the redshift of OII narrow emission line's peak. We start our identification with He I* $\lambda$ 10830. Thanks to its large oscillator strength, it is deeply depleted in a wide range of velocity. Thus, it became easier to identify the other absorption lines, as long as they fall in the same velocity range. Following this lead, we consequentially found the following lines: Ca II $\lambda\lambda$ 3934,3968, He I* $\lambda$ 3889, and Na I $\lambda\lambda$ 5890,5893. Mg II $\lambda\lambda$ 2795,2802 is not clear enough for quantitative analysis, but it is traceable enough to identify. Thus, we also listed it there.

There might be doubt about whether the Ca II and Na I doublets are really coming from the absorbing gas, or are they coming from the stellar radiation. The reason we identify them as absorbing gas originated are as follows. First, they appear to have similar velocity ranges with the He I* $\lambda$ 10830 line, and they are very likely to be coming from the same origin. He I* however, is impossible to be stellar originated. Thus, these two doublets are very likely to have the same origin. Second, we noticed that these lines are around 600 km/s blueshifted to the NEL peak which is usually representative of galactic scale gas, so these two don`t seem to be coming from the same component as NEL is.
Third, this quasar has a rest-frame V-band absolute magnitude of $V\sim-25.1$, much luminous than typical galaxy with $V>-23$, and thus the starlight would contribute only a little in this spectral region and stellar origin cannot explain the observed strong Na ID absorption line.
For all these three reasons, we can identify Ca II and Na I absorption to be coming from a absorbing gas moving several hundred kilometers per second blue-shifted from the galaxy.

\subsection{The covering condition of the absorbing gas}

\subsubsection{solving the equation sets to obtain covering factor of Ca II}

We noticed that Both of CaII doublets are clearly observed and well separated from each other, yet they are close enough in wavelengths to assume they are in the same physical condition(optical covering factor, for example), so we can use the doublet ratio method (Nachmann \& Hobbs 1973) to determine the covering condition of the absorbing gas.

\begin{figure}[!htbp]
\centering
\includegraphics[width=\linewidth]{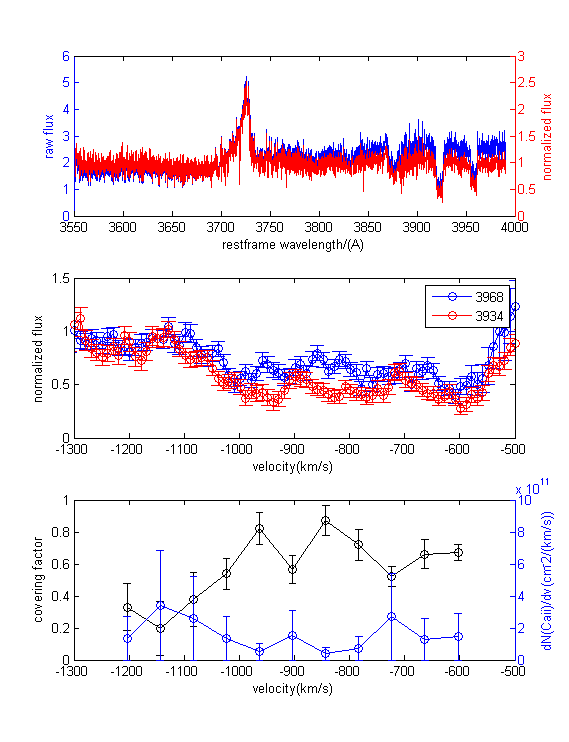}
\caption{Determination of the covering condition. The first column from the top shows the normalization, the second shows the absorption lines in velocity space, while the latter shows the Monte-Carlo result for covering condition.}
\label{cover}
\end{figure}

\normalsize After normalizing the spectrum, we determined the flux distribution with respect to velocity, using the formula
\begin{equation}
\label{cof}
\large f=f_0 (1-cf+cf(1-e^{-\tau}))
\end{equation}
and
\begin{equation}
\label{nt}
\large \frac{dN}{dv}=\frac{m_e c \tau{v}}{\pi e^2 f \lambda}
\end{equation}
\normalsize (Nachmann \& Hobbs 1973), where  f is oscillator strength, which can be obtained from on-the-ground experiments; cf is covering factor, and $\ \frac{dN}{dv}$\  is column density of particles within the range of v~v+dv only dependent on the particle generating absorption lines, disregarding wavelength. Thus, we can solve $\ \frac{dN}{dv}$\ and cf with respect to velocity from the remnant flux of the doublet. And by integrating dN/dv, we got the column density of CaII. In order to increase S/N ratio, we coalesced six data points on the spectrum to get one coalesced point. The best values and uncertainties are calculated using Monte-Carlo methods, and we limited the values in the ranges of $0 < cf < 1$ and $\tau > 0$.
This result is shown in figure \ref{cover}.
We can see from the figures that the covering factor is highest in the velocity range of $-1000<v<-800$ km/s with cf value of 0.7--0.8.
The value is consistent with the lower limit $cf>0.65$ set by the depth of Ca II $\lambda$3934 absorption trough.
The average value of cf is 0.69 in the velocity range of $-1233<v<-573$ km/s, and the integrated column density is $ 1.9\times 10^{14} cm^{-2} $ in this velocity range.

\subsubsection{Fraction of starlight in ESI aperture at this wavelength using HST decomposition result}
\label{hst_analysis}
\begin{figure}[!htbp]
\centering
\includegraphics[width=\linewidth]{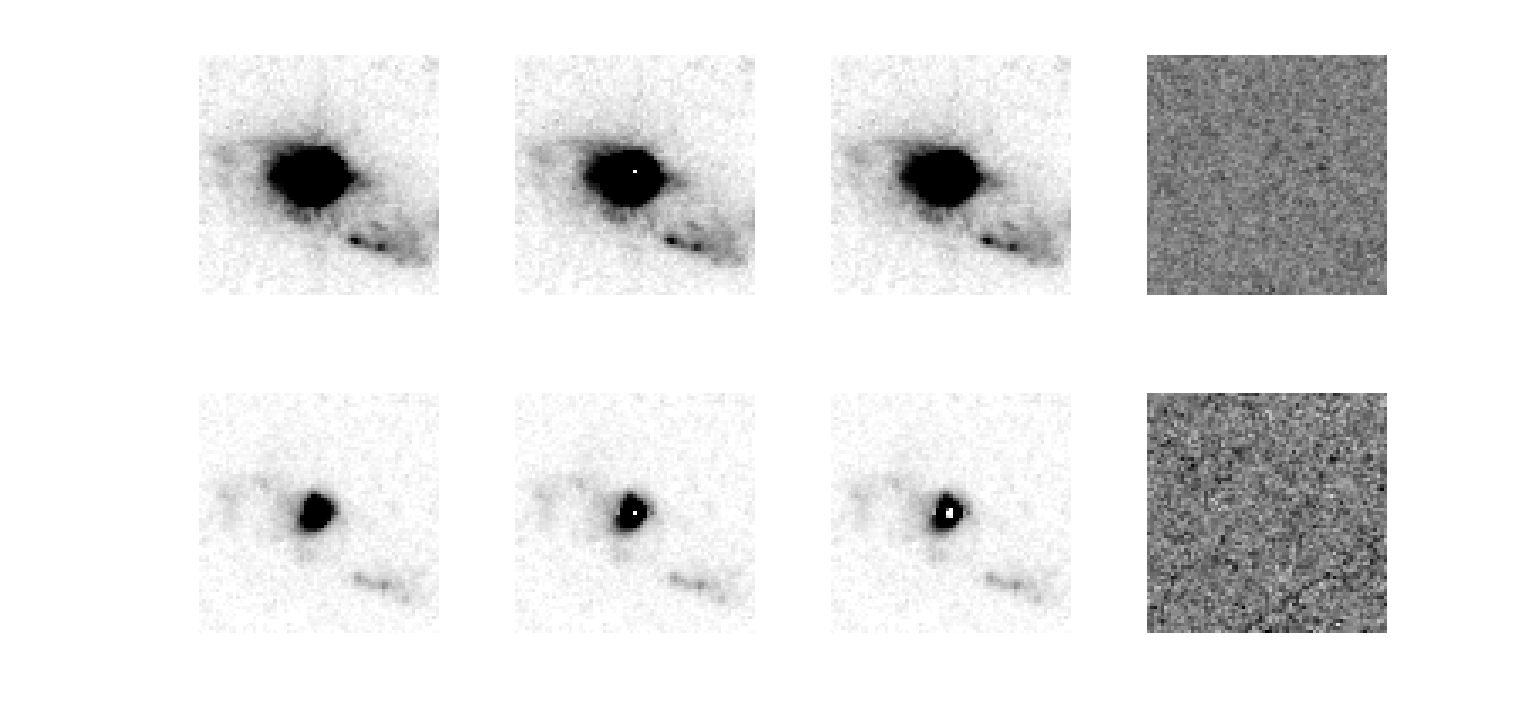}
\caption{HST-ACS image and decomposition. The upper row is the decomposition for F814W, while the lower F475W. In each row, the first subplot on the left is the original image,  2md and 3rd are stellar component derived from "zero" method and GALFIT, respectively, while the last one is the residue.}
\label{hstde}
\end{figure}

By a happy coincidence, we found out that the CaII doublet are located in approximately the same wavelength as HST-ACS F814W band, centered at 8051\AA \ on the earth coordinate, meaning 4121\AA  \  in the rest frame. The ACS image, shown in figure \ref{hstde}, shows that the object is composed of a luminous point source and a spatially resolved extended source.
In order to determine the contribution of the AGN radiation to the total flux, we decomposed the HST-ACS image. We first made a PSF model with a nearby star observation taken from the same day, as stated in Chapter \ref{reduction}. First, we adopted the ¡°zero¡± decomposing method (Urrutia et al. 2008), and conservatively decomposed the image into a point source and a extended source, where the extended source`s luminosity is subtracted to zero at the central pixel. In this way, we can by definition get the upper limit of the point source luminosity. The results are also shown in figure \ref{hstde}. \\

On the other hand, we used an image decomposing software GALFIT, and decomposed the image into PSF and SERSIC components. PSF, as stated above, stands for the quasar luminosity, while SERSIC stands for the galactic radiation. The sersic component is introduced to give a better estimation of the quasar emission; however, it cannot represent the total galactic luminosity. Although the decomposition came back with a $\frac{\chi^{2}}{\nu}=0.510 $ showing that the model roughly fits the data, we do see structures, tidal tails, on the residue image, which for the record is a suggestion for galactic merging (we will further discuss this in Chapter \ref{discussion}). So, in order to obtain more accurate result, we consider the subtraction of the quasar component from the original image, instead of Sersic model, to represent galactic luminosity.\\
In order to make comparisons between ESI data and the images obtained by HST decompositions, we integrated the flux of each component within the spatial range of ESI slit (0.75" $\times$ 3" slit), obtained by both methods. In the zero case, we can see that in F814W, the PSF component flux accounts for 40\% of the luminosity, which can be treated as the upper limit of the fraction of quasar contribution, while the best-estimation of the fraction can be obtained to be 38\% from GALFIT decomposition.

The overall optical covering factor $cf$, obtained from absorption line calculations, is larger than the quasar flux component seen here. If the absorbing gas only covers the quasar, it is obvious that the covering factor is by no means going to be larger than that. Thus, this decomposition is a clear suggestion for at least partial covering of stellar light. In case this quasar is subject to severe variation, we compared the photometry data taken by SDSS in April 2000, which is only 8 months away from the ESI spectrum, to the data taken by HST in October 2005, check for differences in the luminosity. We interpolated the SDSS i-band and z-band magnitudes to 814W band and compared it with HST photometry. It turns out that the variation is only 0.08 magnitude (SDSS photometry is brighter), so it is safe to say the condition stayed stable throughout the five years.
Using the decomposition record above, we can now determine the percentage of stellar radiation covered by the absorbing gas. We assume that the gas has covered all of the quasar luminosity, and part of the stellar luminosity. Through simple algebra, we can get that
\begin{equation}
\label{percent}
 r=1-\frac{L_{uncovered}}{L_{stellar}/L_{total}}=1-\frac{1-0.69}{1-0.38}=49\%
\end{equation}
of the galaxy`s radiation.
With a similar method, we also decomposed the image obtained by F475W, centering at 4745\AA \ (2426\AA \ in rest frame). The result came out as follows: In ``zero'' method, psf component takes up 48\% of the luminosity,  and GALFIT method shows 42\%. The issue here is, however, we get an even bigger quasar component in GALFIT than ``zero'', which violates the definition. Taking into consider that the difference is not big enough to qualitatively affect further conclusions, we simply accept this fact and adapt ``zero'' result in later analysis.

\subsection{ Measuring the ionic column densities}
As mentioned above, we have obtained more than one absorption line in the spectra. These lines and the atoms they stand for can tell us facts about the parameters about the absorbing gas, e.g. depth of the gas, photo-ionization, etc. So, in this part we analyse the other absorption lines. However, because the quasar and host galaxy have different spectra, lines at different wavelengths vary in their optical covering factor. Thus, we first decompose the SED to obtain the ratio of quasar flux to stellar flux at all wavelengths.
\subsubsection{SED fitting to obtain the fraction of starlight at all wavelengths}
\label{sed}
\begin{figure}[!htbp]
\centering
\includegraphics[width=\linewidth]{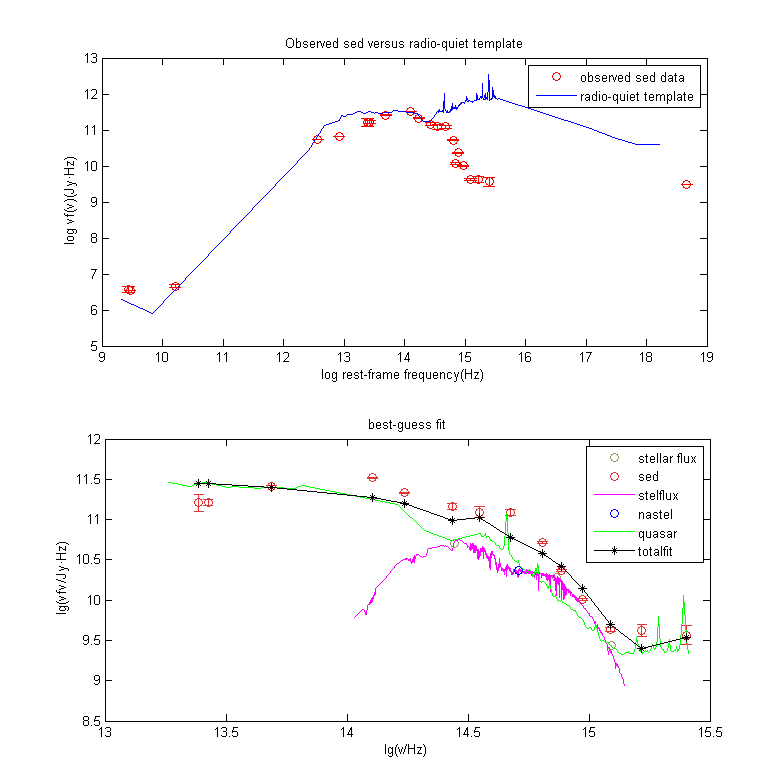}
\caption{SED plot and fitting procedure. Red circles stand for observed SED, black stands for overall fitting in the lower column. The upper figure showed the observed SED plotted against the template, while the lower shows the fitting components.}
\label{sed-fitting}
\end{figure}

As stated in \S \ref{hst_analysis}, the object is made of a quasar and a host galaxy.
The intrinsic quasar SED is more diverse than stellar SED, so we first focus on determining stellar luminosity.
Till now, we have obtained two reliable points where we know the ratio of stellar luminosity versus quasar luminosity: From the decomposition in \S \ref{hst_analysis}, we can get the quasar-to-stellar ratio at F814W, F475W. Furthermore, by assuming the outflow`s optical covering factor for the stellar alone is uniform at all wavelengths, we can also derive the ratio at He I* $\lambda$10830. The reason why we can do this is that, from the absorption lines stated above, we can see obvious absorption of He I* $\lambda$3889 with optical depth $\tau\sim0.4$. He I* $\lambda$10830 has a oscillator strength 23.5 times that of $\lambda$3889, thus it should be nearly depleted considering the high oscillator strength. The covering factor, as a result, approximately equals the fraction of light it observes, which makes its stellar vs total ratio $\frac{L_{stellar}}{L_{stellar}+L_{quasar}}=\frac{L_{stellar}}{L_{total}}=\frac{f_{0}}{r} $ , where r is the part of stellar radiation covered by the outflow, and $f_{0}=0.17$ is the remainder of He I* absorption line. We first took the average covering factor of Ca II $ \lambda \lambda$ 3934,3968, $ cf_{Ca} = 0.69 $ as the total optical covering factor, and assume the quasar has been fully covered on the line of sight. From equation \ref{percent}, we know that r=0.49.  Thus, it is clear that the ratio is
\begin{equation}
 \frac{L_{stellar}}{L_{stellar}+L_{quasar}}=\frac{f_{0}}{r}=0.36
\end{equation}
Then, we can use mark this three wavelengths on the SED figure, and obtain the total luminosity, further stellar luminosity of these points. After that, we try on each spectrum of the Simple Stellar Population(SSP, BC03) templates to fit to the points by reddening with SMC extinction curve, and choose the one with the minimum $\chi^2$ to be our result.
After fitting trial with all the template included in the library, we chose one with metal abundance of $z=0.05=6.25\times z_{sun} $\ and an age of 8.3 Myr. The result also shows that the $E_{B-V}$ for the best fit is 1.20 (corresponding to $A_V=3.3$). We will use this in later discussion.\\
As a bonus, we also fitted the residue radiation, stellar best-fit subtracted from the observation, from the observation collected above. In this part, we assumed that the quasar inside is originally a radio-quiet quasar(as seen evidently in the upper subplot of figure \ref{sed-fitting} )--and thus used the radio-quiet template SED from Shang et al. (2011), and adapted SMC reddening for absorbing gas.
With the same math method as above, we obtained the best-fit parameter of the model. The E(B-V) of the quasar component turned out to be 1.06 (corresponding to $A_V=2.9$), fairly close to the reddening of stellar component, and the intrinsic bolometric luminosity is $7\times10^{46}$ erg/s.
We noticed that the rest-frame UV part of the SED exceeds the reddened quasar template, and accordingly introduced a scattering component, which makes up only 0.28\% of the total luminosity.
The stellar and quasar components are then added up, and shown in the lower subplot of figure \ref{sed-fitting} together with the original observation.

\subsubsection{measurement of column densities}

\begin{enumerate}
\item{He I*}\\
 We have mentioned above that He I* $\lambda$10830 is highly saturated. This is good when determining the covering factor, but will also cause a large uncertainty when calculating the column density. Thus, we use He I* $\lambda$3889 for the column density. The He I* $\lambda$3889 absorption line is close to the Ca II doublet, so it is safe to assume the same covering factor as Ca II doublet. Then, using the same method and dataset of Monte-Carlo simulation as above, we estimated the column density of He I*. The simulation resulted in a column density of $ 6.0\times 10^{14} cm^{-2} $ within the same velocity range as Ca II.
\item{NaI doublet}\\
\normalsize Na I doublet are blended and the covering fraction can not be directly solved.
 Thus, we have to calculate the column density adopting the covering condition determined by \S \ref{sed} and velocity profile obtained by CaII. The oscillating factor for NaI doublets are already given by lab data on earth, so we can directly determine the ratio of the absorption lines` optical depth. With all the assumption above, we can fit the model to the spectrum with only one parameter undetermined, which is the Na I column density. During the fitting process, we broadened the ESI spectrum with a Gaussian distribution, taking $\frac{\lambda}{\delta\lambda}=2000$ as TripleSpec has a lower spectral resolution. The calculation gives a result of $ 1.14\times 10^{14} cm^{-2} $. In this figure, the ``5890'' and ``5896'' lines are drawn as $e^{-\tau_i}$, where $\tau_1$ and $\tau_2$ stands respectively for each wavelength and has a fixed ratio $\frac{0.681}{0.320}$.

\end{enumerate}

\subsection{physical properties of the absorbing gas}
\label{properties}
To this point, we have obtained all the observational parameters we need. Now, we can start to calculate some physical properties of the absorbing gas.
\begin{enumerate}
\item{size of the outflow\\}
  \normalsize We already know that the outflow covers $r=49\% $ of the starlight radiation. To determine the outflow`s size, we get back to figure \ref{hstde}, and look at filter F814W. We found out, from the decomposition image,  that the circle centered at the quasar which covers 49\% percent of the stellar luminosity has a radius of 1.6 kpc and an area of 8 kpc$^2$.
  This is only a lower limit of the crosssectional area of the absorbing gas considering the following:
  \\(a), The surface brightness of a galaxy is higher in the center, thus absorbing gas with other shapes needs more area to cover a certain fraction of starlight.
  \\(b), There may be starlight in front of the absorbing gas.

  \item{H column density\\}
  \normalsize The He I* metastable state is reduced from the recombination of He II, thus is only cumulated in H II zone and stops growing after ionization front.
  On the other hand, Ca II and Na I can only exist in H I zone.
  According to Liu15, when the ionization front is well developed, the column density of He I* is roughly a single-valued function with respect to U, which is defined as:
  \begin{equation}
  \large U=\frac{Q_{H}}{4 \pi R^{2}n_{H} c}
  \end{equation}
  Given that we have seen Na ID absorption, this gas is bound to have a very deep H I zone.
  Using the results of Liu15, the ionization parameter is $\ U=10^{-1.6}=0.025$, and the column density of the H II zone is $2\times10^{21}$ cm$^{-2}$.
  \\The depth of H I zone can be obtained from column densities of Na I and Ca II.
  This need three correction: element abundance, ionization state and dust depletion.
  We used the solar abundance.
  The correction for ionization is more complex and we used photo-ionization code Cloudy (version 10.00, Ferland et al. 1998) to simulate the ionization process in the absorbing gas.
  We assumed a slab-shaped geometry, unique density, homogeneous chemical composition of solar values, and an SED of ionizing continuum of commonly used MF87 (Mathews \& Ferland 1987).
  We also assumed that the gas is homogeneously mixed with dust and the dust to gas ratio is set to be same as SMC.
  The ionization parameter $U$ are set to be $10^{-1.6}$.
  We tried several Hydrogen densities and found that the results are independent of density, and thus set the density to be the typical ISM density of $10^3$ cm$^{-3}$.
  The simulation demonstrated that the fraction of $N_{Na I} / N_{Na}$ increases with the depth, being $\sim10^{-3}$ at ionization front and $\sim10^{-1}$ at a place with $N_H=3\times10^{22}$ cm$^{-2}$.
  By adopting a Na dust depletions of $10^{-1}$, the measured Na I column density requires a total H column density of $3\times10^{22}$ cm$^{-2}$, resulting an averaged $N_{Na I} / N_{Na}$ fraction of $\sim$3\%.
  If assuming a Ca depletion of $10^{-2}$, observed Ca II requires a similar value, and the averaged $N_{Ca II} / N_{Ca}$ fraction is $\sim$40 \%.
  The results from Na I and Ca II are consistent with each other, and thus we obtained the best estimation of $\sim3\times10^{22}$ cm$^{-2}$.
  \\On the other hand, the column density of H I zone can be estimated from dust reddening.
  The $E_{B-V}$ of stellar and AGN components are 1.20 and 1.06 from the SED fitting result if assuming SMC extinction curve.
  Since the reddening of stellar component represents the most part of the absorbing gas, $E_{B-V}$ value of 1.2 was used.
  Bouchet et al. (1985) gave the gas-to-dust ratio of SMC, i.e., \\
  \begin{equation}
    \large  R=\frac{N_{H}}{E(B-V)}=3.7 \  to \  5.2 \times 10^{22}
  \end{equation}
  Thus we derive that\\
  \begin{equation}\label{NH}
  \large N_{H}= 4.4\ to \ 6.2\times 10^{22} cm^{-2}
  \end{equation}
  This value are close to that from Na I and Ca II, supporting the result from absorption lines.
  In addition, H column density can also be obtained from X-ray data.
  However, the X-ray spectrum J0729 has too few ($\sim$6) photons (Urrutia et al. 2005) to do so.

  \item{Mass, dynamic timescale and outflow rate\\}
  \normalsize Assuming hydrogen abundence X=75\%, we get $M = \frac{Area \times N_H}{m_{H}\times X}>3 \times 10^9 M_\odot.$ \\
  \normalsize Dynamic timescale $\ t=\frac{R}{v}=2 Myr $\, where v is the mean velocity of the N-v velocity contour of 800 km/s. \\
  \normalsize The outflow rate equals $\frac{M}{t}> 1.5\times 10^3 M_\odot$/yr .

  \item{Density of the gas\\}
  we can derive
  \begin{equation}\label{nH}
  \large n_{H}=\frac {Q_{H}}{4 \pi R^{2} U c}
  \end{equation}
  \normalsize Now let`s first determine $\ Q_{H}$\ . At this point, we assumed that the inner redden-free quasar has the same SED as an average radio-quiet quasar. We fit the seemingly redden-free radio-FIR part to the synthesized spectrum, with one simple scaling factor, which turned out to be 11.46.\\
  \normalsize Then, we can get
  \begin{equation}
  \large Q_{H}=\int_{\nu_{0}}^{\inf} \frac {f_{\nu } 4 \pi d_{L}^{2}}{h \nu }d\nu = 9 \times 10^{56} s^{-1},
  \end{equation}
  where $\ d_{L} = 6.274 Gpc $\ .
  The distance of the absorbing gas to the central ionizing source is unknown.
  Assuming the distance is the same with the size 1.6 kpc, we get $n_H=4\times10^3$ cm$^{-3}$.
  Although having large uncertainty due to uncertainties in $Q(H)$ and $R$, this value is consistent with the density of galactic-scale ISM of $10^2$--$10^4$ cm$^{-3}$.
  The averaged depth of the absorption gas is $\ D= N_{H} \div n_{H} \sim 3 pc $\ .

\end{enumerate}

\section{Discussion}
\label{discussion}
\subsection{Massive Galactic-scale Outflow}

From the parameters and calculation above, we can clearly see that this object contains a very large-scaled outflow. As stated above, the absorbing gas has a few blue-shifted absorption lines, covers a large portion of the luminosity, and has a size comparable to the whole galaxy. From both the SED and the spectra, we are very confident that we have discovered direct evidence for galactic-scaled outflow. And, from figure \ref{lines}, we can see that these outflows extend to a velocity up to -1300 km/s, which is very unusual for starburst-driven outflows. However, this is quite normal among AGN-driven outflows. Thus, we are confident enough to say that we have spotted evidence for a large-scaled, AGN driven outflow. Just for the record, we have also seen that the blue wing of the OII emission line is significantly larger than the red wing, which is also consistent with this conclusion. \\

The conclusion on the outflow's size relies on the result that the absorbing gas covered 49\% of stellar radiation(see equation \ref{percent}). We now inspect the reliability of this result.
This number is obtained by the fact that the absorbing gas covered 69\% of the continuum in Ca II H\&K doublet region, and that the HST/ACS image show that quasar only takes up 38\% of the continuum in F814W band.
The usage of Ca II doublet in determining covering condition is long-tested, and the high quality of ESI spectrum yields a 10\% accuracy of the covering factor, which is from a Monto-Carlo simulation.
Also, the depth of Ca II $\lambda$3934 absorption line gives an independent lower limit of $\sim$65\%.
On the other hand, the fraction of quasar emission is also robust.
The zero decomposition provides a reliable upper limit of 40\% in ACS image.
Although the rest-frame wavelengths of F814W band (4130 \AA) and Ca II doublet (3950 \AA) are slightly different,  the fractions of quasar emission in these two wavelengths would be close because the SED fitting result shows that the slope of SED are close for quasar and stellar components in $\lambda\sim4000$ \AA\ (Figure 5).
As we have showed, the variability between spectral and imaging observations is only 0.08 magnitude, and the change in fraction of quasar emission would be only 5\%, which can neither explain the difference.
Therefore it is safe to say that the absorbing gas covers a significant fraction of starlight, which leads to the conclusion that the gas is galactic scale.

When measuring the total mass of the absorbing gas, there are two basic parameters, the crosssectional area and the H column density.
The crosssectional area was from the fraction of starlight which the absorbing gas covers.
As we have pointed out in Section 3.3, the area calculated by assuming a circle geometry is only a lower limit.
Furthermore, considering the possible contourpart of this absorbing gas in the opposite direction, the total mass of galactic outflow may be multipled by a factor of 2 assuming a bipolar cone-like geometry, or a factor of 4 assuming a sphere-like geometry.

The H column density is calculated via two method, from column densities of Na I and Ca II, and from dust to gas ratio.
Conversion of Na I and Ca II rely on element abundance, ionization state and dust depletion.
Here we discuss the main uncertainties.
\\1. The metallicity of ISM increases with increasing stellar luminosity in general. The host galaxy of J0729 is very bright thus the abundance of Na and Ca in the absorbing gas may be higher than solar values.
However, this will not significantly affect the estimation of the H column density.
We noticed that in the Cloudy simulation the accumulated $N_{Na I}/N_{Na}$ is very sensitive to $N_H$ in the range of $N_H(IF) < N_H < 10 \times N_H(IF)$, where $N_H(IF)$ is the column density in the ionization front.
In other words, the $N_H$ is insensitive to $N_{Na I}/N_{Na}$.
For example, if the abundance of Na is 1 dex higher than the solar one, which is a rather extreme assumption, the inferred H column density is only 0.5 dex lower.
\\2. $N_{Na I}/N_{Na}$ is dependent on the ionization parameter $U$ and the amount of dust mixed in the gas.
The robustness of calculating $U$ from He I* column density is testified by Liu et al. (2015).
We attempted to add more dust in the absorbing gas than we have assumed (SMC), and found that the fraction of Na I would be higher, but the difference is small.
\\3. In the simulation we assumed a Na dust depletion of $10^{-1}$ and Ca depletion of $10^{-2}$.
Observed values in the MW is typically lower than these values.
If adopting lower values, the H column density would be higher.
\\Besides conversion from Na I and Ca II, there is independent estimation of $N_H$ from reddening measurement and dust to gas ratio, which is in good agreement.
In addition, the 0.2--10 keV X-ray flux from Chandra (Urrutia et al. 2005) is about one order of magnitude lower than the predicted value by assuming the intrinsic SED of J0729 is the same as averaged quasar SED, which is consistent with H column density between $10^{22}$ and $10^{23}$ cm$^{-2}$.
In summary, the estimation of $N_H\sim3\times10^{22}$ cm$^{-2}$ is trustworthy and it is likely that $N_H>10^{22}$ cm$^{-2}$ even under extreme assumptions.

We have presented a massive ($M>3\times10^9 M_\odot$) galactic scale outflow from the analysis of absorption lines.
A question is that what drives this outflow, AGN or starburst?
Rupke et al. (2005) investigated neutral gas outflow using Na ID absorption line in a sample of starburst dominated ULIRGs, and found that the averaged velocity of the absorption trough is no more than $-600$ km/s.
Thus it is likely that the neutral gas outflow with $-800$ km/s in J0729 is driven by AGN or by both AGN and starburst.
Although there are some galaxies host high-velocity outflows without AGN (e.g. Diamond-Stanic et al. 2012), they are typically UV-luminous and unobscured, not similar to J0729.

The neutral gas outflow seen in Ca II and Na I absorption may be not the complete picture of the outflow in J0729.
The [O II] $\lambda$3727 emission line host a strong blue wing which extents to $-3000$ km/s (see Figure 3), indicating an ionized gas outflow with higher velocity.
The outflow at velocity $>1200$ km/s is not seen in Ca\&Na absorption trough, suggesting that the correponding gas may be too thin to developed an H I zone.
Considering the higher outflowing velocity, we guess that the ionized gas outflow may be in an inner region and to be the upstream of the neutral gas outflow.

\subsection{The Scenario}

\normalsize The massive outflow seen in J0729 is undoubtly rare and interesting. A reader might ask what caused such a phenomenon. Here is our best-thought somehow na\"{i}ve guess of the physical process concerning this object.

\normalsize From figure \ref{hstde}, we can see that this galaxy has a tidal tail in the lower-right region of the image, as well as the upper-left region. This is a clear trace for galactic interaction.
Together with other pieces of evidence, we are able to make the guess that this object recently go a major merger process
which results in the quasar we see.
The collision of two gas-rich galaxies triggered both SMBH accretion and starburst, and the later produced large amount of dust which accounted for the high reddening.
As the AGN began to shine, it either sublimated the dust in the inner core, blows out the dust in the outer region by radiation pressure, or got rid of it by heat carried by energetic outflow, after which the AGN core began to show its face.
However, given the limited time, the incomplete process leaves us a heavily reddened quasar.

This model is able to explain many phenomenons we have observed.
The dusty-rich gas in the host galaxy, accelerated by the AGN or starburst, account for the blueshifted absorption line in the spectrum.
The dynamic time scale of the outflow is $\sim$2 Myr, in agreement with that the AGN was recently born and have not yet digested all the dust.
Also, although with large uncertainty, the SED fitting result (Figure 5) show that the stellar component is dominated by a young population, supporting that J0729 hosts recent starburst.

In this study, we combined spatial image decomposition and absorption line analysis to identify a massive galactic scale outflow in a reddened quasar J0729.
Furthermore, the total mass of the outflowing gas can be estimated from the covering factor of the absorbing gas and ionic column densities.
If the ULIRG--quasar evolutionary model is correct, one can find more such massive galactic scale outflows from reddened quasars.
Therefore we suggest a systematic search of this type of outflow in reddened quasars by taking high resolution spectral and imaging observations.
This may help the study of massive galactic-scale outflow and AGN negative feedback.\\
\ \\

\centerline{REFERENCE\\}
\normalsize \ \\
Sturm, E.; Gonzalez-Alfonso,E.; Veilleux,S.; et al, 2011, ApJL, 733, L16\\
Sanders, D. B.; Soifer, B. T.; Elias, J. H.; et al., 1988, ApJ, 325, 74\\
Hopkins, P. F., Murray, N.; Thompson, T. A.; 2009, MNRAS, 398, 303\\
Ferrarese,L.; Merritt,D.; 2000, ApJ,539L,9F\\
Di Matteo, T.; Springel, V.; Hernquist, L.; 2005, Nature, 433, 604\\
Veilleux, S.; Cecil, G.; Bland-Hawthorn, J. 2005, ARA\&A, 43, 769\\
Spoon,H.W.W.; Farrah, D.; Lebouteiller, V.; 2013,ApJ,775,127S\\
Cicone, C.; Maiolino, R.; Sturm, E.; 2014,A\&A,562A,21C\\ 
Wilson, John C.; Henderson,Charles P.; Herter, Terry L.; et al., 2004, SPIE, 5492, 1295W\\
Cushing, Michael C.; Vacca, William D.; Rayner, John T.; 2004, PASP, 116, 362C\\
Nachman, P.; Hobbs, L.M.; 1973, ApJ, 182, 481N\\
Urrutia,T.; Lacy,M.;Becker,R.H.;2008, ApJ, 674, 80\\
Shang, Zhaohui; Brotherton, Michael S.; Wills, Beverley J.; et al., 2011ApJS, 196, 2S\\
Liu,W.; Zhou,H.; Ji,T.; et al., arXiv 1502.00240, Accepted by ApJS\\
Ferland, G. J.; Korista, K. T.; Verner, D. A.; et al., 1998, PASP, 110, 761F\\
Mathews, William G.; Ferland, Gary J., 1987, ApJ, 323, 456M\\
Urrutia, Tanya; Lacy, Mark; Gregg, Michael D.; Becker, Robert H., 2005, ApJ, 627, 75U\\
Rupke, David S.; Veilleux, Sylvain; Sanders, D. B.; 2005, ApJS, 160, 87R\\
Diamond-Stanic, Aleksandar M.; Moustakas, John; Tremonti, Christy A.; 2012, ApJ, 755L, 26D\\

\end{document}